\documentclass[10pt]{article}

\usepackage{amsmath,amssymb,mathrsfs,graphicx}
\usepackage{nature_djp}

\title{Entangling two atoms of different isotopes via Rydberg blockade}

\author{Y. Zeng$^{1,2,3}$, P. Xu$^{1,2}$, X.D. He$^{1,2}$, Y.Y. Liu$^{1,2,3}$, M. Liu$^{1,2}$, J. Wang$^{1,2}$,  D.J. Papoular$^{4}$,
  G.V. Shlyapnikov$^{5,6,7,8}$, M.S. Zhan$^{1,2}$ }

\usepackage{bm}

\renewcommand{\Re}{\operatorname{Re}}

\begin{document}

%\linenumbers

\maketitle

\begin{affiliations}
 \item State Key Laboratory of Magnetic Resonance and Atomic and Molecular Physics, Wuhan Institute of Physics and Mathematics, Chinese Academy of Sciences - Wuhan National Laboratory for Optoelectronics, Wuhan 430071, China
 \item Center for Cold Atom Physics, Chinese Academy of Sciences, Wuhan 430071, China
 \item University of Chinese Academy of Sciences, Beijing 100049, China
 \item LPTM, UMR8089 of CNRS and Univ.~Cergy--Pontoise,
   F--95302 Cergy--Pontoise, France
\item LPTMS, UMR8626 of CNRS and Univ.~Paris--Sud,
  F--91405 Orsay, France
\item SPEC, CEA \& CNRS, Univ.~Paris--Saclay, CEA Saclay,
  F--91191 Gif--sur--Yvette, France
\item Russian Quantum Center, Novaya Street, Skolkovo,
  Moscow Region R--143025, Russia
\item Van der Waals--Zeeman Institute, Institute of Physics,
   Univ.~Amsterdam, The Netherlands
\end{affiliations}

\begin{abstract}
Quantum entanglement is crucial for simulating and understanding exotic physics of strongly correlated many-body systems, such as high--temperature superconductors, or fractional  quantum   Hall  states
\cite{laflorencie:PhysRep2016,sachdev:PTRSA2016}. The entanglement of non-identical particles exhibits richer physics of strong many-body correlations \cite{sachdev:PTRSA2016,amico:RMP2008} and offers more opportunities for quantum computation \cite{Saffman-review2016}, especially with neutral atoms where in contrast to ions the interparticle interaction is widely tunable by Feshbach resonances \cite{chin:RMP2010}. Moreover, the inter-species entanglement forms a basis for the properties of various compound systems \cite{tichy2011}, ranging from Bose-Bose mixtures \cite{wang2016} to photosynthetic light-harvesting complexes \cite{sarovar2010}.
So far, the inter-species entanglement has only been obtained for trapped
ions\cite{ion-1,ion-2}.
Here we report on the experimental realization of entanglement of two
neutral atoms of different isotopes. A ${}^{87}\mathrm{Rb}$ atom and a
${}^{85}\mathrm{Rb}$ atom are  confined in  two  single--atom  optical
traps separated by 3.8 $\mu$m \cite{Peng2015}.
Creating a strong Rydberg blockade, we demonstrate  a  heteronuclear
controlled--NOT  (C--NOT)  quantum  gate and generate a  heteronuclear entangled state, with
raw fidelities $0.73  \pm 0.01$ and $0.59 \pm 0.03$, respectively.
Our work, together with the technologies of single--qubit gate and C--NOT gate developed for
identical atoms, can be used for simulating any many--body system with multi-species interactions.
It also has applications in quantum computing and quantum metrology, since heteronuclear systems exhibit advantages in low crosstalk and in memory protection.
\end{abstract}

Trapped neutral atoms
offer  unique
possibilities for quantum simulation,
thanks to an excellent control of the interaction strength
over 12 orders of magnitude\cite{Saffman_RMP}  and to the creation of
tunable  arrays of  single atoms  for the
simulation of  spin systems\cite{Antoine2016}.  Important
experiments have been
performed towards quantum simulation
using identical neutral atoms\cite{JILA-collision,
Ryd-dressing,2D-addressing,3D-addressing,array-1,array-2},
and theoretical proposals aim at universal
simulators\cite{Weimer-Rydberg-simulator}.
Mixed--species architectures further enlarge the set of systems that can
be simulated to encompass new phenomena
ranging from heteronuclear Efimov effects\cite{ulmanis:PRL2016}
to exotic superfluid pairing mechanisms in quantum fluid
mixtures\cite{rysti:PRB2012,ferrierbarbut:Science2014}.

Heteronuclear qubits are also helpful
for solving fundamental issues in quantum information
processing, such as low--crosstalk
individual manipulation \cite{Saffman-review2016}.
The two different Rubidium isotopes used in our experiment
exhibit different level structures,
allowing us to implement a  novel technique to
manipulate individual
atom states by using a difference between the transition
frequencies of the two atoms \cite{ion-2}.
This feature is a fundamental difference
compared to previous experiments on identical
atoms \cite{Antoine-entanglement,Saffman-CNOT},
where individual addressing relied on the spatial separation
between the atoms. In our setup, the atoms do not need to be spatially
separated, and all laser beams cover both atoms. We implement this technique
for the first time with neutral atoms and show that
the fidelities of the created CNOT quantum gate and entangled state
are on par with recent
homonuclear results \cite{Antoine-entanglement,Saffman-CNOT}.
Our analysis shows that the  fidelities
are mainly limited by technical reasons and by the thermal motion of the atoms.

\begin{figure}
\includegraphics[width=\linewidth]{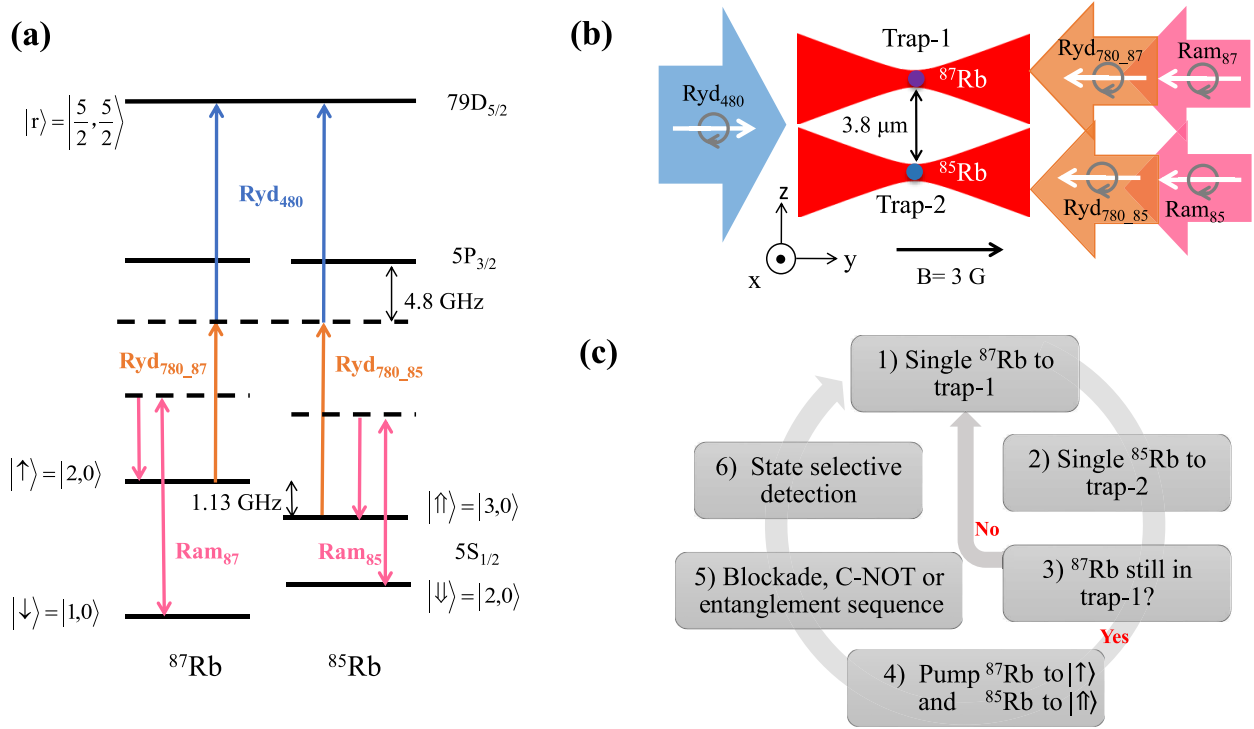}
\caption{\label{fig:expsetup}  Experimental Setup.  (a) Energy  levels
  and lasers for $^{87}\mathrm{Rb}$ and
  $^{85}\mathrm{Rb}$.
  Atoms are excited to
  Rydberg  states through
  Raman transitions  using
  $480\,\mathrm{nm}$
  ($\mathrm{Ryd}_{480}$)  and
  $780\,\mathrm{nm}$  ($\mathrm{Ryd}_{780}$)
  $\sigma^+$--polarised lasers.  The
  laser
  $\mathrm{Ryd}_{480}$  is
  blue--detuned
  by  $4.8\,\mathrm{GHz}$  from the  intermediate state, and
  its waist $12.8\,\mu\mathrm{m}$ covers both atoms.
  The lasers $\mathrm{Ryd}_{780-87}$ and $\mathrm{Ryd}_{780-85}$,
  whose frequencies differ by $1.13\,\mathrm{GHz}$,
  address $^{87}\mathrm{Rb}$ and $^{85}\mathrm{Rb}$.
  The
  degeneracy of  the Rydberg states $|79D_{5/2},m_j\rangle$  is lifted
  by the static magnetic field $B=3\,\mathrm{G}$
  along the quantization axis $y$,
  and the laser frequencies are resonant with
  the $m_j=5/2$ state.  Single qubit operations are performed through
  Raman  transitions using the $795\,\mathrm{nm}$ lasers
  $\mathrm{Ram}_{85}$  and  $\mathrm{Ram}_{87}$,
  which are red--detuned by
  $50\,\mathrm{GHz}$  from  the
  $5S_{1/2}\rightarrow5P_{1/2}$    transition.
  (b)    Experimental
  geometry. Two  $830\,\mathrm{nm}$ lasers
  have the beam waist $2.1\,\mu\mathrm{m}$ to form
  two dipole traps separated by $3.8\,\mathrm{\mu m}$ along the $z$ direction.
  (c) Experimental time  sequence.}
  \end{figure}

In our experiment, we fully control two heteronuclear atom qubits
represented by a single ${}^{87}\mathrm{Rb}$ atom and a single
${}^{85}\mathrm{Rb}$ atom, and exploit the heteronuclear Rydberg
interaction to deterministically
entangle  the two  different atoms.   The control qubit is encoded
in the  ground  hyperfine
states
$|F=1, m_F=0\rangle=|\downarrow\rangle$              and
$|2,0\rangle= |\uparrow  \rangle$ of  $^{87}$Rb,
whereas  the   target   qubit   is encoded in   the   states
$|2,0\rangle=|\Downarrow\rangle$ and $|3,0\rangle=|\Uparrow\rangle$ of
$^{85}$Rb (Fig.~1a).    For both atoms, the     Rydberg    state     is
$|r\rangle=|79D_{5/2},m_j=5/2\rangle$.
We exploit the difference in the resonance frequencies of the two
atoms to ensure a negligible crosstalk during state measurements
and qubit operations (see Methods).

The  experimental apparatus  and the single--atom trapping procedure
for $^{87}$Rb and $^{85}$Rb  atoms have been described in our
recent  work\cite{Peng2015}.
We trap a single  $^{87}$Rb  atom in  the dipole  trap--1
and a  single $^{85}$Rb  atom in the dipole  trap--2 located $3.8\,\mathrm{\mu  m}$  away  (see Fig.~1b),
and then optically pump the atoms
to the  $|\uparrow\rangle$  and $|\Uparrow\rangle$  states,
respectively. After that the trapping potentials are adiabatically lowered from 0.6 mK to  0.1mK.
Both microtraps have trapping frequencies
$\omega_y/2\pi=1.39\pm 0.01\,\mathrm{kHz}$ in the longitudinal direction
and
$\omega_r/2\pi=16.9\pm 0.1 \,\mathrm{kHz}$ in the radial direction
(see Fig.~\ref{fig:expsetup}b).
We measure the atom temperatures
$T_{87}=8 \pm 1 \, \mu\mathrm{K}$
and  $T_{85}=9\pm 1 \mu\mathrm{K}$  using
release and recapture methods.
Next, we combine Rydberg
excitation pulses  and single qubit  operations with Raman  lasers
in order
to  demonstrate the  heteronuclear  Rydberg
blockade, implement the C--NOT  gate, and entangle the two
heteronuclear
atoms.
At the end of each sequence, we detect the qubit state by using
a   resonant    laser   to   ``blow   away"    $|\uparrow\rangle$   and
$|\Uparrow\rangle$   atoms,  so that    the   survival
probabilities  refer  to the  atoms  in  the $|\downarrow\rangle$  and
$|\Downarrow\rangle$ states (see Fig.~\ref{fig:expsetup}c).

We  first calculate the expected Rydberg blockade  shift.
If both atoms are  in the $|r\rangle$
state, their  interaction is dominated  by the F\"orster  resonance
between   the   two--atom   states  in   the   $(79d_{5/2},79d_{5/2})$,
$(80p_{3/2},78f)$,    and   $(81p_{3/2},77f)$    manifolds.
We   restrict  the  F\"orster
interaction  Hamiltonian   to  a   subspace  spanned  by   436  states
corresponding  to distinguishable atoms.
Taking the initial state $|r\!\Uparrow\!>$ we account for its coupling to
the F{\"o}rster states and
calculate the
time evolution   of    the   probability    for   double excitation,
$P_{85}(y,t)=1-|<\!r\!\Uparrow|e^{-iHt/\hbar}|r\!\Uparrow\!>|^2$   and
its  average  over time,  $P_{85}(y)$.  The latter depends on  the  offset
$y=|y_2-y_1|$ of  the two  atoms along  the $y$  direction.
The   blockade   shift\cite{Saffman2009}
$\Delta E(y)$  is deduced from the relation
$P_{85}(y)=(\hbar\Omega_{85})^2/((\hbar\Omega_{85})^2+{\Delta}E^2)$,
where $\Omega_{85}$ is the effective Rabi frequency for ${}^{85}\mathrm{Rb}$.
At zero temperature, for the distance $z=3.8~\mathrm{\mu  m}$  between the microtraps,
assuming a spatial offset of $y=1\,\mu\mathrm{m}$,
the effective Rydberg interaction between the atoms is close to the strongly--interacting
F\"orster regime\cite{Saffman_RMP}.
Accordingly, the numerical results
yield $P_{85}\approx 10^{-6}$ and a very large blockade shift
$\Delta E/h=600\,\mathrm{MHz}$
(see   Methods  and  Supplemental   Material).
The finite  temperature of the atoms causes  them to explore
larger values  of the  offset, $y\gtrsim  10\,\mathrm{\mu m}$,
leading  to  the  mean
double--excitation  probability  $<P_{85}>\approx   0.013$  for  our
temperatures $T_{87}=8\,\mu$K and $T_{85}=9\,\mu$K.

\begin{figure}
\includegraphics[width=\linewidth]{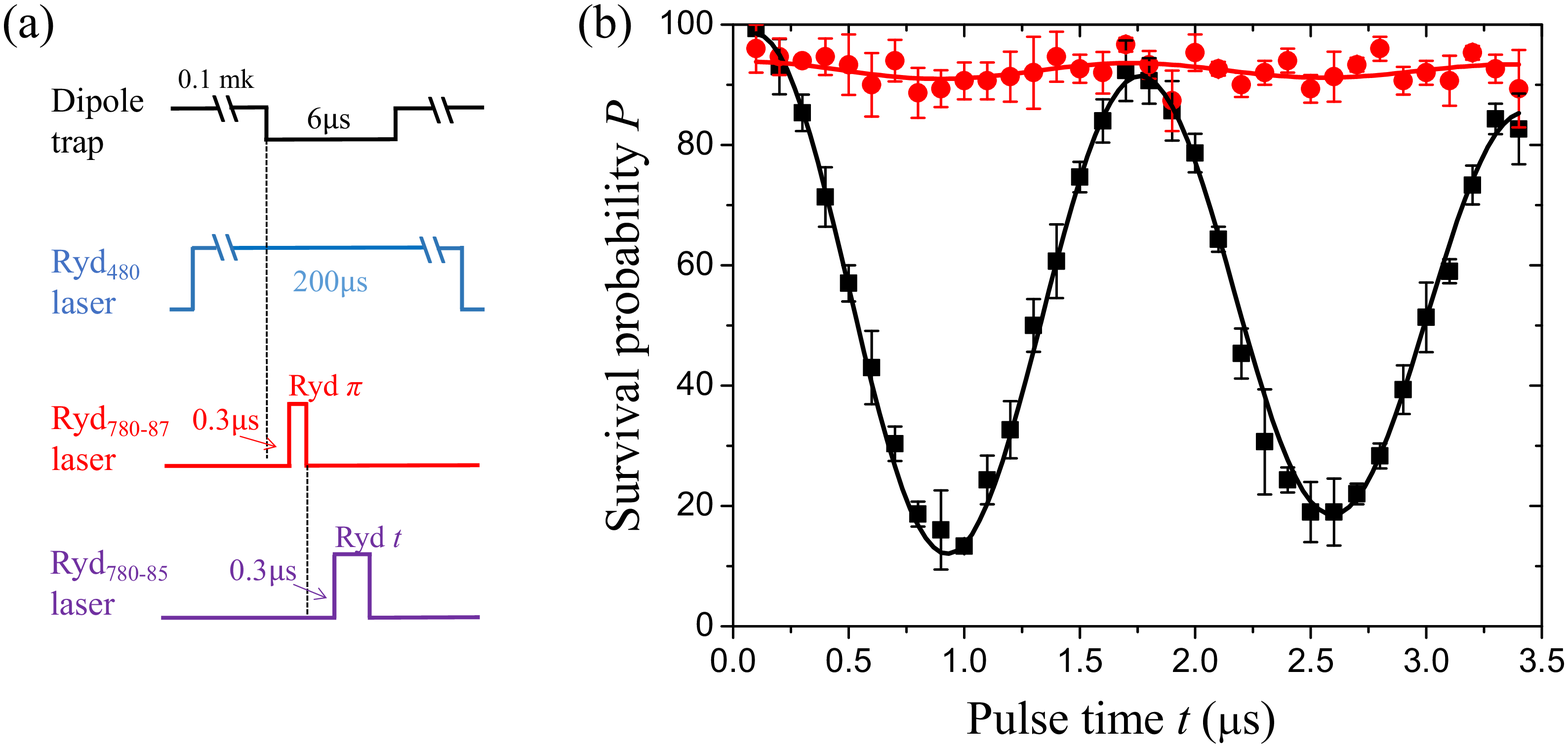}
\caption{\label{fig:rydbergblockade-exp}     Heteronuclear     Rydberg
  blockade.  (a) Time sequence.
  (b) Rabi  oscillations between  the
  $^{85}\mathrm{Rb}$ $|\Uparrow\rangle$  and  $|r\rangle$   states.
  The experimental data are shown both in the absence (black squares)
  and in the presence (red circles)
  of $^{87}\mathrm{Rb}$ in  trap--1.
  The  solid curves are damped sinusoidal fits
  with peak--to--peak amplitudes
  $0.91 \pm 0.02$ (black squares) and
  $0.03 \pm 0.01$ (red circles). Each data point represents the average
  over $150$ repetitions, and the error bars correspond to one standard deviation.}
\end{figure}

We demonstrate the Rydberg blockade by
applying a Rydberg $\pi$ pulse on $^{87}$Rb,
waiting  for 0.3 $\mu$s,
and  applying a Rydberg pulse of variable duration
on ${}^{85}\mathrm{Rb}$ (Fig.~2a). We
measure the Rabi oscillations
between the  $^{85}$Rb  $|\Uparrow\rangle$ and $|r\rangle$ states as a function of the second pulse duration (Fig.~2b).
The Rydberg states are detected through the atom  loss
with an efficiency
of $\sim$90\%, and the Rydberg  excitation efficiency for  $^{87}$Rb and $^{85}$Rb is  $\sim$96\% (see Methods).
The lifetime of the $|r\rangle$ state is over  180 $\mu$s,
providing a long enough blockade for $^{85}$Rb.
We  do not record the experimental data
when $^{87}$Rb is still in the trap after the sequence,
so as
to eliminate unblockaded events when $^{87}$Rb is not excited to the $|r\rangle$ state.
The peak  to peak amplitude
of  $^{85}$Rb Rabi  oscillations  between  the $|\Uparrow\rangle$  and
$|r\rangle$ states  is 0.91 $\pm$0.02  in the absence of $^{87}$Rb
in trap--1 (Fig.~2b). In its presence, the experimental data show a strong Rydberg  blockade which
suppresses the  oscillation amplitude to 0.03  $\pm$0.01,
in accordance with our theoretical prediction.  The remaining
weak  oscillations  of  $^{85}$Rb  are   mainly  due  to not perfect
experimental  conditions,   including  the   loss  of   $^{87}$Rb  and
transitions to other Rydberg states.

We use the Rydberg blockade to generate a
heteronuclear C--NOT gate
following the protocol of Ref.\cite{Jaksch2000}.
This involves three Rydberg pulses (Fig.~3a):
\textit{(i)} a   $\pi$ pulse on $^{87}$Rb between the $|\uparrow\rangle$  and  $|r\rangle$ states, \textit{(ii)} a $2\pi$ pulse on $^{85}$Rb between $|\Uparrow\rangle$ and  $|r\rangle$, and \textit{(iii)} a $\pi$ pulse on $^{87}$Rb between $|r\rangle$ and $|\uparrow\rangle$. Then, combining two Hadamard gates realized using Raman $\pi/2$ pulses between  the $|\Uparrow\rangle$ and $|\Downarrow\rangle$ states, we demonstrate the heteronuclear C--NOT gate shown in Fig.~3.
Its intrinsic coherence is illustrated
by measuring the oscillation of the output probabilities
as  a function of the relative  phase between  the two Hadamard  gates
(Fig.~3b).
Setting the
relative phase to 0 ($\pi$), the C--NOT gate will flip the target qubit
if   the   control   qubit   is   $|\uparrow\rangle$
($|\downarrow\rangle$).

\begin{figure}
\includegraphics[width=\linewidth]{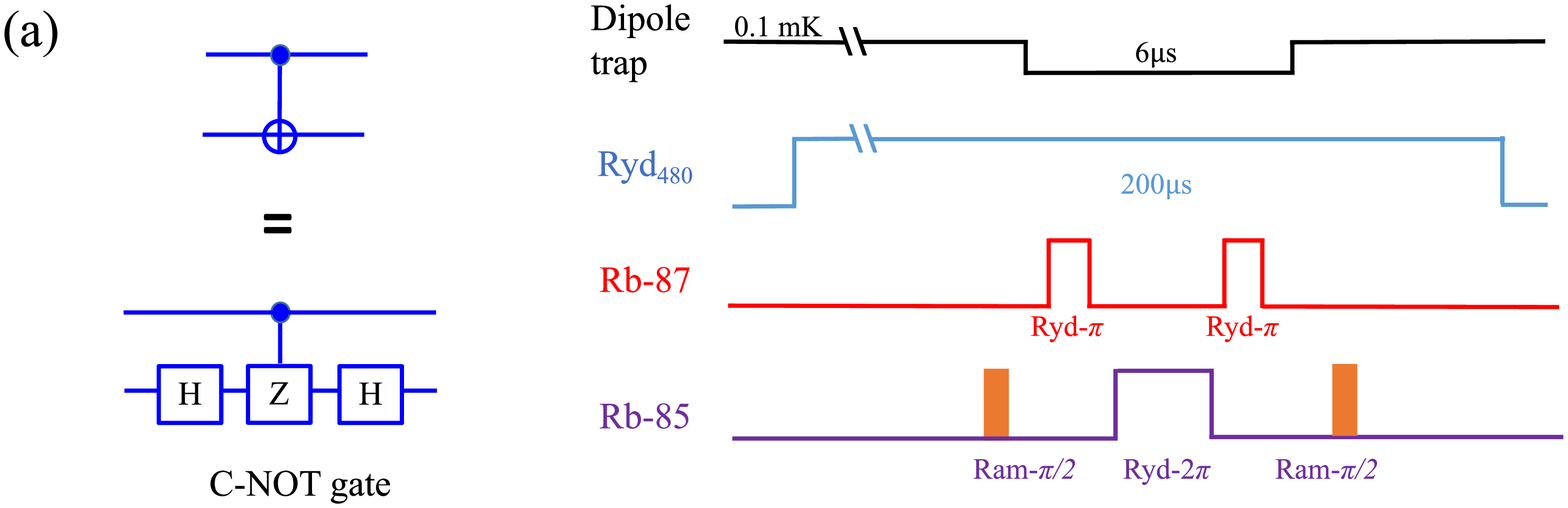}
\includegraphics[width=\linewidth]{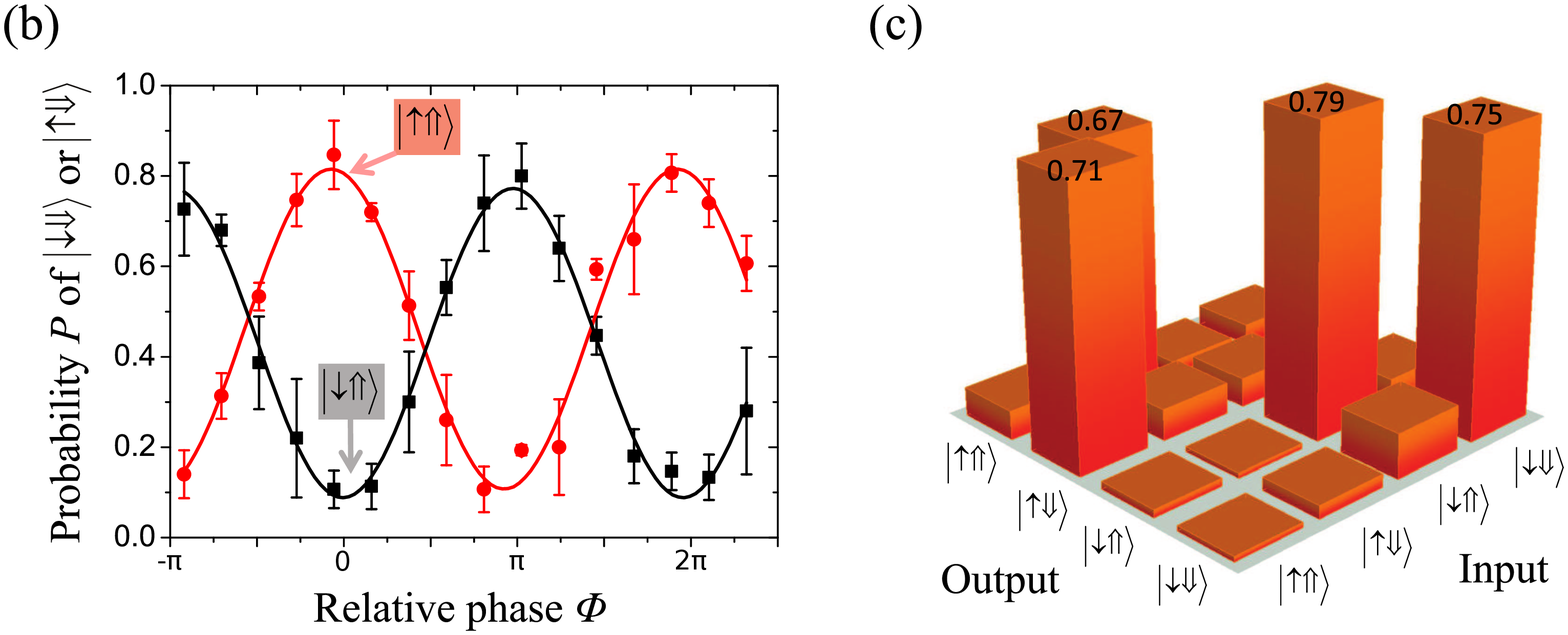}
\caption{Heteronuclear C--NOT gate.
  (a) Experimental time sequence.
  (b) Output  states
  as a function of the relative phase  between the Raman
  $\pi/2$  pulses, for the initial states
  $|\downarrow\Uparrow\rangle$ (black) and
  $|\uparrow\Uparrow\rangle$ (red).
  The  solid  curves are  sinusoidal  fits
  yielding the phase difference
  of $(0.94 \pm 0.01)\pi$ between the two signals.
  (c) Measured probability matrix $U_\mathrm{CNOT}$ for
  the C--NOT gate with the relative phase between the
  $\pi/2$ pulses set to 0.
  Each data point represents $150$ repetitions, and the error bars correspond to one single deviation.}
\end{figure}

The fidelity of the  CNOT gate is determined by measuring its
truth table probabilities (Fig.~3c). We add an
extra Raman $\pi$ pulse before acting with the ``blow away" laser
to transfer the $|\uparrow\rangle$ state ${}^{87}\mathrm{Rb}$ atoms
to  $|\downarrow\rangle$
and the $|\Uparrow\rangle$ state ${}^{85}\,\mathrm{Rb}$
atoms
to  $|\Downarrow\rangle$, in order to exclude
other atom    losses    as in
Ref.\cite{Saffman-CNOT}.
The      raw     fidelity      of      the      C--NOT     gate      is
$F=\operatorname{Tr}[|U_\mathrm{ideal}^{T}|U_\mathrm{CNOT}]/4=0.73(1)$.
It is mainly limited by technical reasons and can be made higher by stabilizing the Raman pulse powers
and by improving the Rydberg excitation efficiency (see Methods).

Eventually,  we deterministically
generate a heteronuclear entangled  state of
${}^{87}\mathrm{Rb}$ and ${}^{85}\mathrm{Rb}$.
Starting with the   two--atom state
$(|\uparrow\rangle+i|\downarrow\rangle)|\Downarrow\rangle/\sqrt{2}$,
we  apply  the  C--NOT gate  to  create the
entangled state
$(|\uparrow\Uparrow\rangle+|\downarrow\Downarrow\rangle)/\sqrt{2}$.
In order to quantify the  entanglement of our created Bell state,
we measure the coherence $C$ between the $|\uparrow\Uparrow\rangle$
and  $|\downarrow\Downarrow\rangle$ states
by  studying   the  response
of  the system to  the  simultaneous rotation  of  the
two qubits\cite{turchette:PRL1998}.
For that purpose, we apply to both atoms $\pi/2$ pulses
carrying  the same phase $\phi_1$ relative to the initial pulses (Fig.~4a)
and  measure the oscillations of the parity  signal
$P=P_{\uparrow\Uparrow} +P_{\downarrow\Downarrow}
  -P_{\uparrow\Downarrow}-P_{\downarrow\Uparrow}$  as a
function of $\phi_1$ (Fig. 4c).
This gives us access\cite{turchette:PRL1998,Antoine-entanglement}
to the coherence $|C|=0.16\pm 0.01$
which, combined with the populations $P_{\uparrow\Uparrow}=0.41$
and $P_{\downarrow\Downarrow}=0.44$ (Fig. 4b),
leads to the entangled state fidelity
$F=
  (P_{\uparrow\Uparrow}+P_{\downarrow\Downarrow})/2
  +|C|=
  0.59\pm0.03$.
The obtained fidelity is clearly above the threshold of $0.5$ ensuring
the presence of entanglement.
We obtain it
without any  corrections for atom or trace losses.
It  is  lower  than the fidelity  of  our
C--NOT gate
mainly because of the motion
of $^{87}$Rb. Following Ref.\cite{Antoine-entanglement}
we evaluate that at our temperatures and C-NOT gate fidelity
the upper bound of the entanglement fidelity
is $F_\mathrm{ent-max}= 0.65$, which
is   slightly    above   our   experimental result.

\begin{figure}
\includegraphics[width=0.95\linewidth]{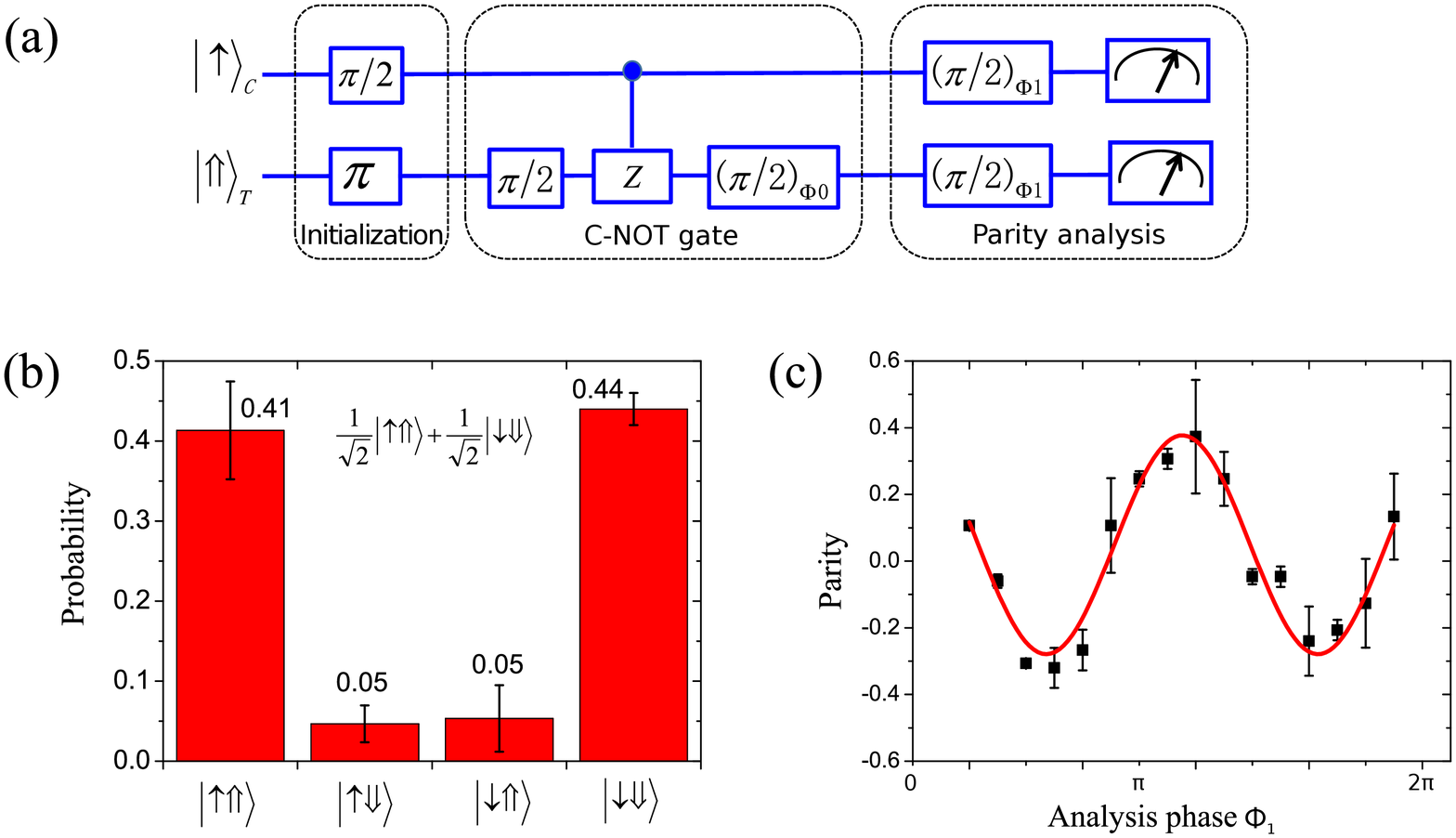}
\caption{Deterministic entanglement of two  heteronuclear atoms.
  (a) Time  sequence. (b)
  Measured  probabilities   for  the entangled   state.
  (c)   The  parity
  signal $P$.    The   solid    curve   is    a sinusoidal   fit    with
  $P=2\Re(C_{2})-2|C_{1}|\cos(2\phi_{1}+\xi)$,
  where $\Re(C_{2})=0.02 \pm 0.02$, $|C_{1}|= 0.16 \pm 0.01$.
  Each data point represents $150$ repetitions, and the error bars correspond to one standard deviation.}
\end{figure}

To conclude,  we have  realized a C--NOT  gate between  two
non--identical
single atoms and  demonstrated a negligible crosstalk  between the two
atomic qubits.  The gate  is based on  a strong  heteronuclear Rydberg
blockade, and the raw fidelity is 0.73 $\pm$ 0.01. The entanglement of
two different atoms is  then deterministically generated with
the raw fidelity
$0.59 \pm  0.03$.
Our   work  makes  a  significant   step  towards  the manipulation of
heteronuclear atom systems. Unlike identical atoms, we use a difference in
the transition frequencies to individually address a single atom. In this case, the two
atoms can be put at a short separation while maintaining individual addressing
to explore the physics in a very strong Rydberg interaction regime.
Many atoms representing different isotopes can be trapped
in an array with an arbitrary geometry\cite{array-1,array-2}
to  realize a  Rydberg  quantum
simulator of  exotic spin models,  such as the Kitaev toric code,  color code,
or coherent  energy  transfer.
Our results pave a way towards quantum computing
with heteronuclear systems %\cite{Saffman-review2016}
and  towards the realization of  a  high fidelity  state  detection,
which has recently been predicted not to have any fundamental limit even
at room temperature\cite{HigherFidelity}.

%% Put the bibliography here, most people will use BiBTeX in
%% which case the environment below should be replaced with
%% the \bibliography{} command.

\begin{addendum}
 \item[Acknowledgements] This work was supported by the National Key Research and Development Program of China under Grants No.~2016YFA0302800, the National Natural Science Foundation of China under Grants No.~11674361, the Strategic Priority Research Program of the Chinese Academy of Sciences under Grant No.~XDB21010100 and Youth Innovation Promotion Association CAS No.~2017378.
 GVS acknowledges support from IFRAF. DJP and GVS emphasize that the research leading to their results in this paper has received funding from the European Research Council under European Community's Seventh Framework Programme (FR7/2007-2013 Grant Agreement no.~341197).

 \item[Author Contributions]
Y.Z. and P.X. contributed equally to this paper;
Y.Z., P.X., X.D.H., Y.Y.L., M.L., J.W. and M.S.Z. designed and performed
the experiment and analyzed the experimental data;
D.J.P. and G.V.S. performed the theoretical modelling and numerical
calculations;
P.X., D.J.P., G.V.S., and M.S.Z wrote the manuscript; All authors discussed the manuscript.
M.S.Z. supervised the project.
 \item[Competing Interests] The authors declare that they have no competing financial interests.
 \item[Correspondence]
   Correspondence and requests for materials should be addressed to
   P.X.~(email: etherxp@wipm.ac.cn) and M.S.Z.~(email: mszhan@wipm.ac.cn)
\end{addendum}

\begin{methods}
In the first paragraph, we describe the laser system used in our
experiment to realize a coherent Rydberg excitation.
In the second paragraph, we describe our procedure demonstrating
that the crosstalk between the control and target qubits is negligible.
The third paragraph is dedicated to the discussion of the C-NOT gate
and entanglement fidelities.
Finally, in the fourth paragraph, we summarize our calculations
of the Rydberg blockade shift, which is described in greater detail
in the Supplemental Material.

\subsection{Laser system for coherent Rydberg excitations}
The narrow linewidth and stabilized laser source required to
realize a  coherent Rydberg excitation
are  challenging to set up.
In our experiment, the Ryd$_{480}$ Ryderg laser is  generated from
a TA--SHG pro  with a seed laser whose wavelength is 960 nm.
The frequencies of the lasers Ryd$_{480}$ and  Ryd$_{780}$ are locked
to a Fabry--Perot cavity with high finesse (58000 for
$960\,\mathrm{nm}$ and 91000 for $780\,\mathrm{nm})$.
We then reduce the linewidth  to $\sim$10 kHz
for Ryd$_{780}$  and to  $\sim$20 kHz for  Ryd$_{480}$. The  long--term
drift  of both  lasers is  less than  50 kHz.   The
frequency of  Ryd$_{480}$ is set  to $625253.6\,\mathrm{GHz}$,
and we
expand the beam waist of Ryd$_{480}$ to $\sim12.8\,\mathrm{\mu m}$,
so that it covers both atoms. The Ryd$_{780}$  laser light is
divided  into  two beams  with  the  frequency difference
$1127\,\mathrm{MHz}$,
corresponding to the difference in the excitation frequencies of
${}^{85}$Rb (Ryd$_{780-85}$)  and  ${}^{87}$Rb (Ryd$_{780-87}$).  The
frequencies of the Ryd$_{780-87}$ and Ryd$_{780-85}$  lasers
are   $384223.2  \mathrm{GHz}$   and $384224.3 \,\mathrm{GHz}$, respectively. The beam waist of Ryd$_{780-87}$ laser is $\sim7.1\,\mathrm{\mu m}$ , and Ryd$_{780-85}$ laser has the beam waist of $\sim7.8\,\mathrm{\mu m}$.
We use PID controllers with holding function to lock the laser power of Ryd$_{480}$ to 51 mW, and the power of Ryd$_{780-87}$ and Ryd$_{780-85}$ to 5.6 $\mathrm{\mu W}$.
The pulse area
fluctuations of the Ryd$_{480}$ and Ryd$_{780}$ laser pulses are suppressed  to  less  than 1\%. Using the method from Ref.\cite{Saffman2009}, we estimate $\Omega_{780-87}=2\pi\cdot$   226   MHz, $\Omega_{780-87}=2\pi\cdot$   206   MHz, and $\Omega_{480-85}$= $\Omega_{480-87}$= 2$\pi\cdot$  28   MHz;

Coherent Rabi oscillations between the ${}^{85}\mathrm{Rb}$
$|\Uparrow\rangle$  and $|r\rangle$ states and between the ${}^{87}\mathrm{Rb}$
$|\uparrow\rangle$ and $|r\rangle$ states are shown in
Fig.~2b and in Extended Data Fig.~1b.
For $^{87}$Rb, the peak to  peak Rabi amplitude
is 0.82 $\pm$ 0.02.
The  survival  probability  of  $^{87}$Rb   after  a  $\pi$  pulse  is
13\%.   This    includes   the    4\%   probability    of   populating
the $|\uparrow\rangle$ state, the rest being the result of
spontaneous emission from the
Rydberg  state during  the  detection.  Thus,  the Rydberg  excitation
efficiency for  $^{87}$Rb is  $\sim$96\% and the  detection efficiency
for the Rydberg  state is $\sim$90\%. The corresponding efficiency
for $^{85}$Rb  is almost the same.

\subsection{Crosstalk}
The crosstalk of the two atomic qubits
is crucial for our setup because  all lasers cover both atoms, and the
individual  addressing  of  a  single atom  relies  on  the
difference  between the resonance frequencies of $^{87}$Rb  and  $^{85}$Rb
rather  than  on the  spatial
distribution.   During   qubit  state   measurements,  the
$^{85}$Rb resonant laser may cause unwanted scattering of $^{87}$Rb as
it is  detuned 1.1GHz from its resonance frequency,
and  vice versa. We  check this
influence by adding a $^{85}$Rb ``blow away"  pulse between the
$^{87}$Rb ground state Rabi oscillation  and the $^{87}$Rb ``blow away"
pulse. We  then compare  the Rabi oscillations  of $^{87}$Rb  with and
without the $^{85}$Rb pulse as shown in Extended Data Fig.~1a.
The amplitudes of
the Rabi  oscillations are  equal to each other within the  measurement uncertainty,
which  shows a  negligible  crosstalk in  the  state measurement.  For
the excitation to Rydberg states,
we  use two--photon transitions  with the total
Rabi frequency of  about 1MHz. Thus, the GHz  spectral difference can
provide enough  protection for the  qubit operations with  each single
atom. We  also observe almost  no excitation of $^{85}$Rb  when adding
the $^{87}$Rb Rydberg  excitation laser as  shown in Extended Data Fig.~1b.
All experimental data show a negligible  crosstalk between the two  atomic qubits, which represents an important advantage of heteronuclear atom systems.

\begin{figure}
\includegraphics[width=.5\linewidth]{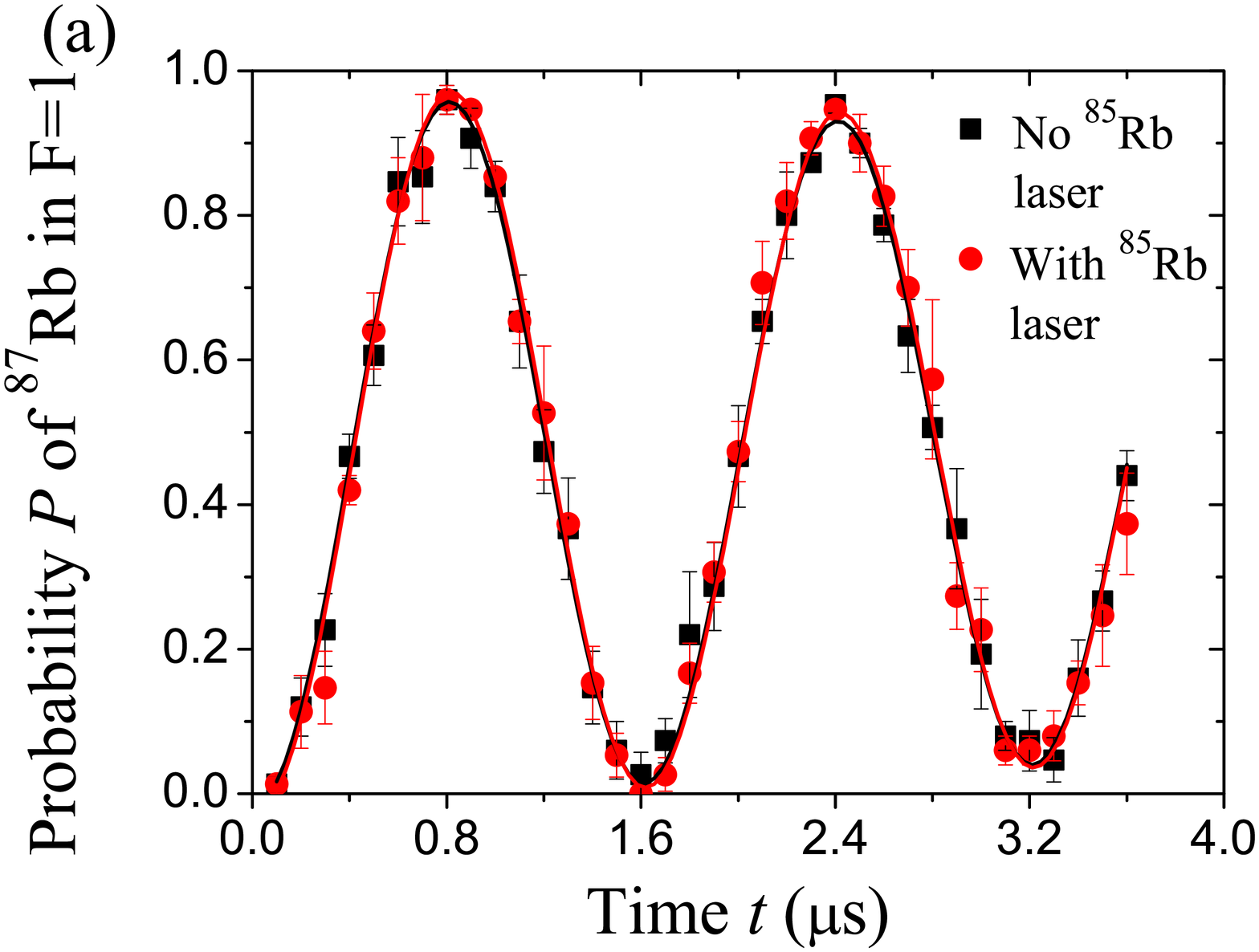}
\includegraphics[width=.5\linewidth]{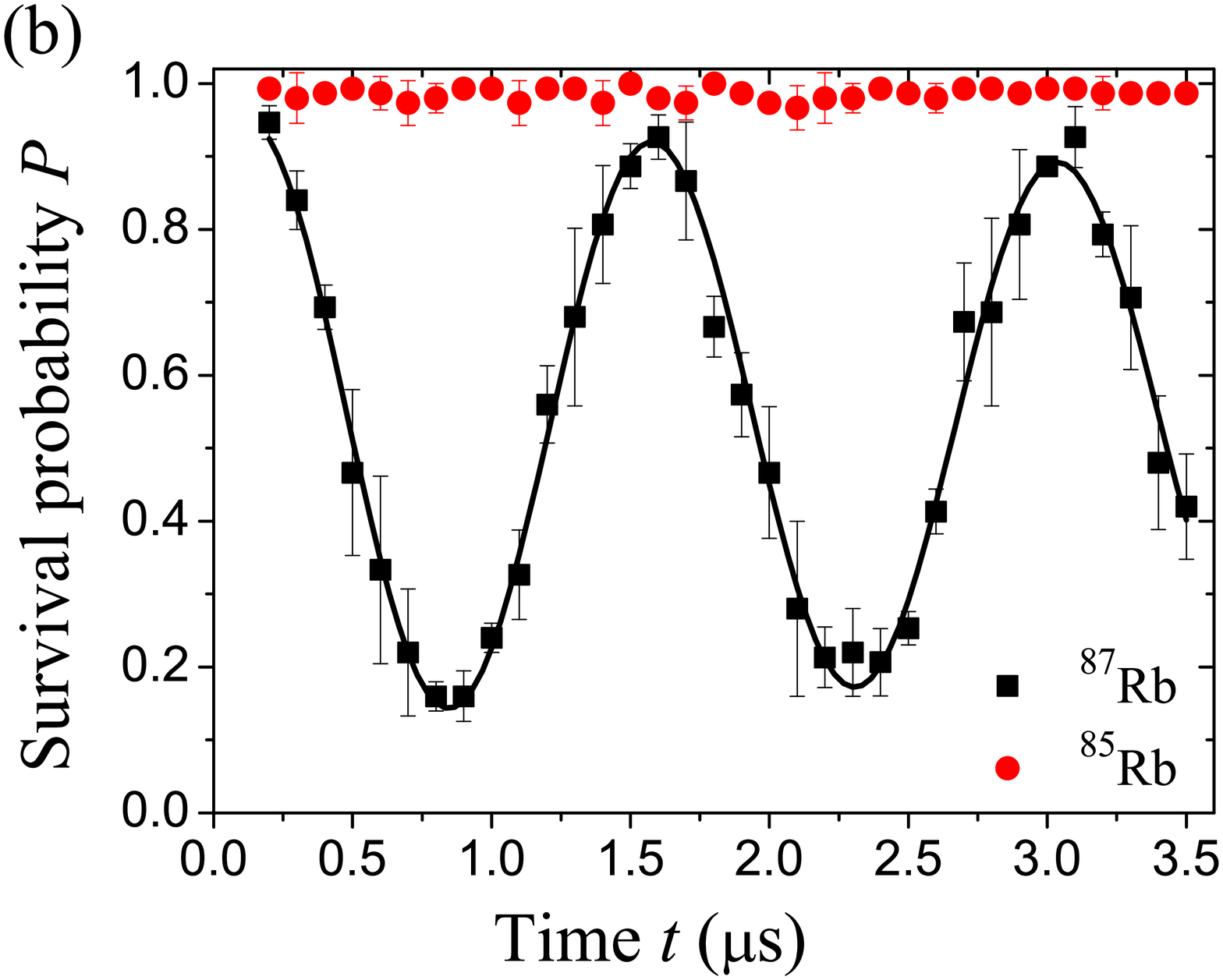}
%\caption {
\textbf{Extended Data Figure 1 }
  Crosstalk   between  $^{85}$Rb   and   $^{87}$Rb.  (a)   Rabi
  oscillations between the $^{87}$Rb
  $|\uparrow\rangle$ and $|\downarrow\rangle$
  states   of  $^{87}$Rb   (black  squares).   The  red   circles  show
  the experimental data obtained
  when  using the $^{85}$Rb ``blow  away'' laser before
  measuring  the  state of  $^{87}$Rb.  The  solid curves  are  damped
  sinusoidal                       fits
  $P=P_0+Ae^{-t/t_0}\cos(2\pi f \,(t-t_c))$, with
  $A=0.49\pm 0.01$, $f=  0.625\pm 0.002\,\mathrm{MHHz}$,
  and $t_0=  28\pm  7\, \mu\mathrm{s}$ for black  squares
  and  $A=0.50  \pm0.02$,
  $f= 0.625\pm 0.003 \,\mathrm{MHz}$,
  $t_0= 27\pm  15\,\mu\mathrm{s}$
  for red circles. (b)
  The $^{87}$Rb Rydberg excitation laser covers both
  $^{87}$Rb in trap--1
  (black squares) and $^{85}$Rb  in trap--2 (red circles). The
  $^{87}$Rb atom
  shows coherent Rabi oscillations  between the $|\uparrow\rangle$ and
  $|r\rangle$ states. The solid  curves are damped sinusoidal fits
  with
  $A=0.41    \pm 0.01$,     $f=   0.685\pm    0.008\,\mathrm{MHz}$,
  and $t_0= 19\pm 5\,  \mu\mathrm{s}$.
  The $^{85}$Rb atom is almost unaffected,
  which shows a negligible crosstalk.
%}
\end{figure}

%**********Heteronuclear Rydberg blockade******

\subsection{The CNOT gate and entanglement fidelity}
The fidelity of the CNOT gate is mainly limited  by technical reasons. One of them is the long--term drift ($\sim$10\%) of the Raman pulse powers, which  reduces the accuracy of Raman  $\pi/2$ and $\pi$ pulses and causes
the fidelity loss of $\sim$9\%. Another reason is the $\sim$96\% Rydberg excitation  efficiency, which causes  about  12\% of  two--atom loss.  By using the power stabilization to suppress the long term drift and employing the compensating stray electrical field to improve the Rydberg excitation efficiency \cite{Ryd-dressing}, one should get a significantly higher fidelity of the C--NOT gate.

The  fidelity  of the  entangled  state  is  lower  than that  of  the
heteronuclear C--NOT gate. This is  mainly due to the motion
of the $^{87}$Rb atom \cite{Antoine-entanglement}. Single  $^{87}$Rb atoms
accumulate  stochastic  phases
$\Phi=\textbf{k}\cdot  \textbf{v} \delta  t$
during the time $\delta t$ separating two  Rydberg--${\pi}$  pulses.
Here,
$|\textbf{k}|= 2\pi/\lambda_{480}-2\pi/\lambda_{780}$,
and $\textbf{v}$ is
the atomic velocity.
These phases vary from shot to shot.
A simple estimation of the average yields
$\langle    e^{i\Phi}\rangle
  = e^{-\langle\Phi^2\rangle/2}
  = e^{-T|\textbf{k}|^2\delta t^2/m_{87}}$,
where  $m_{87}$  is  the  mass  of  $^{87}$Rb,
and we took into account that $\langle{\bf v}^2\rangle=2T/m_{87}$.
With  $T_{87}=  8  \mu  K$  and  $\delta  t=  3.6  \mu  s$,
we find $\langle  e^{i\Phi}\rangle  =0.78$,
implying a  maximum  fidelity  of
$F_{\langle  e^{i\Phi}\rangle}  =  0.89$.  We combine this value
with  the  C--NOT
fidelity to obtain
the maximum entanglement  fidelity $F_\mathrm{ent-max}= 0.65$, which
is   slightly    above   the    upper limit   of    our   experimental
result. According to this calculation, an increase of the fidelity
will  rely on decreasing  the  time gap  between two  Rydberg--${\pi}$
pulses by  increasing the  intensity of  the lasers,  and on lowering
the atom temperatures by using adiabatic cooling.

\subsection{Calculation of the heteronuclear Rydberg blockade shift}

Our theoretical model for the Rydberg blockade involves three steps,
detailed in the Supplemental Material.
First, we characterise the F{\"o}rster resonance
assuming that both atoms are immobile.
The two--atom interaction Hamiltonian $H_F$
accounts for the dipole--dipole interaction between the atoms, the Rydberg energy defects,
and the Zeeman interaction of each atom with the static magnetic field.
If both atoms are excited to Rydberg states with energies close
to the $79d_{5/2}$ state, their interaction is dominated by
the F\"orster resonance\cite{Saffman_RMP}
involving the states in the
$(79d_{5/2},79d_{5/2})$, $(80p_{3/2},78f)$, and $(81p_{3/2},77f)$
manifolds. This amounts to restricting $H_F$
to a subspace spanned by 436 states.

Second, we calculate the blockade shift for fixed atoms(see Extended Data Fig.2a).
Taking the initial two--atom state $|r\!\Uparrow\!>$ one sees that it is coupled to the
doubly--excited F{\"o}rster states  via the operator $W$ which is
proportional to $\Omega_{85}$. The probability for
finding the atom pair in a doubly--excited state after a Rydberg pulse
of duration $t$ on ${}^{85}\mathrm{Rb}$ is
$P_{85}(y,t)=1-|<\!r\!\Uparrow|e^{-iHt/\hbar}|r\!\Uparrow\!>|^2$, where
$H=H_F+W$ and the offset is $y=|y_2-y_1|$.
We numerically calculate its average over time, $P_{85}(y)$.
Following Ref.\cite{Saffman2009}, we then define the blockade shift as
$\Delta E(y)=\hbar\Omega_{85}(1/P_{85}(y)-1)^{1/2}$.
For $y\lesssim 1\,\mathrm{\mu m}$, we find very large
blockade shifts $\Delta E/h\gtrsim 600\,\mathrm{MHz}$ due to a strong
F{\"o}rster resonance with an effective interaction scaling as $1/R^3$, where $R$ is the internuclear distance. The blockade
shift decreases for larger offsets and is of the order of a few
$\mathrm{MHz}$
for $y\gtrsim 10\,\mathrm{\mu m}$.

Finally, we evaluate the impact of finite temperatures
by averaging $P_{85}(y)$ over the probability density
$p(y)$ for the atoms to have the offset $y=|y_2-y_1|$ (see Extended Data Fig.2b).
In the conditions of our experiment the motion of atoms is classical, and
$p(y)$ is Gaussian with the standard deviation
%$\sigma=[k_B(T_{87}/m_{87}+T_{85}/m_{85})/\omega_y^2]^{1/2}$.
$\sigma=[k_BT/(m_\mathrm{red}\omega_y^2)]^{1/2}$,
where the reduced mass is $m_\mathrm{red}=m_{87}m_{85}/(m_{87}+m_{85})$,
and the average temperature $T$ satisfies the relation
$T/m_\mathrm{red}=T_{87}/m_{87}+T_{85}/m_{85}$.
The temperatures and trapping frequencies only
enter  our   model  through  the  combination   $T/\omega_y^2$,  which
characterises the spatial extent of  the classical motion of the atoms
along $y$.
For the experimental values $T_{87}=8\,\mathrm{\mu K}$,
$T_{85}=9\,\mathrm{\mu K}$,
and $\omega_y/2\pi=1.39\,\mathrm{kHz}$, we find
the average value  $<\!P_{85}\!>\approx 0.013$,
which is of the same order  as the observed
quenched Rabi oscillation amplitude (Fig.\ref{fig:rydbergblockade-exp}b).

\begin{figure}
    \begin{minipage}{.5\textwidth}
      \includegraphics[angle=-90,width=\textwidth]
      {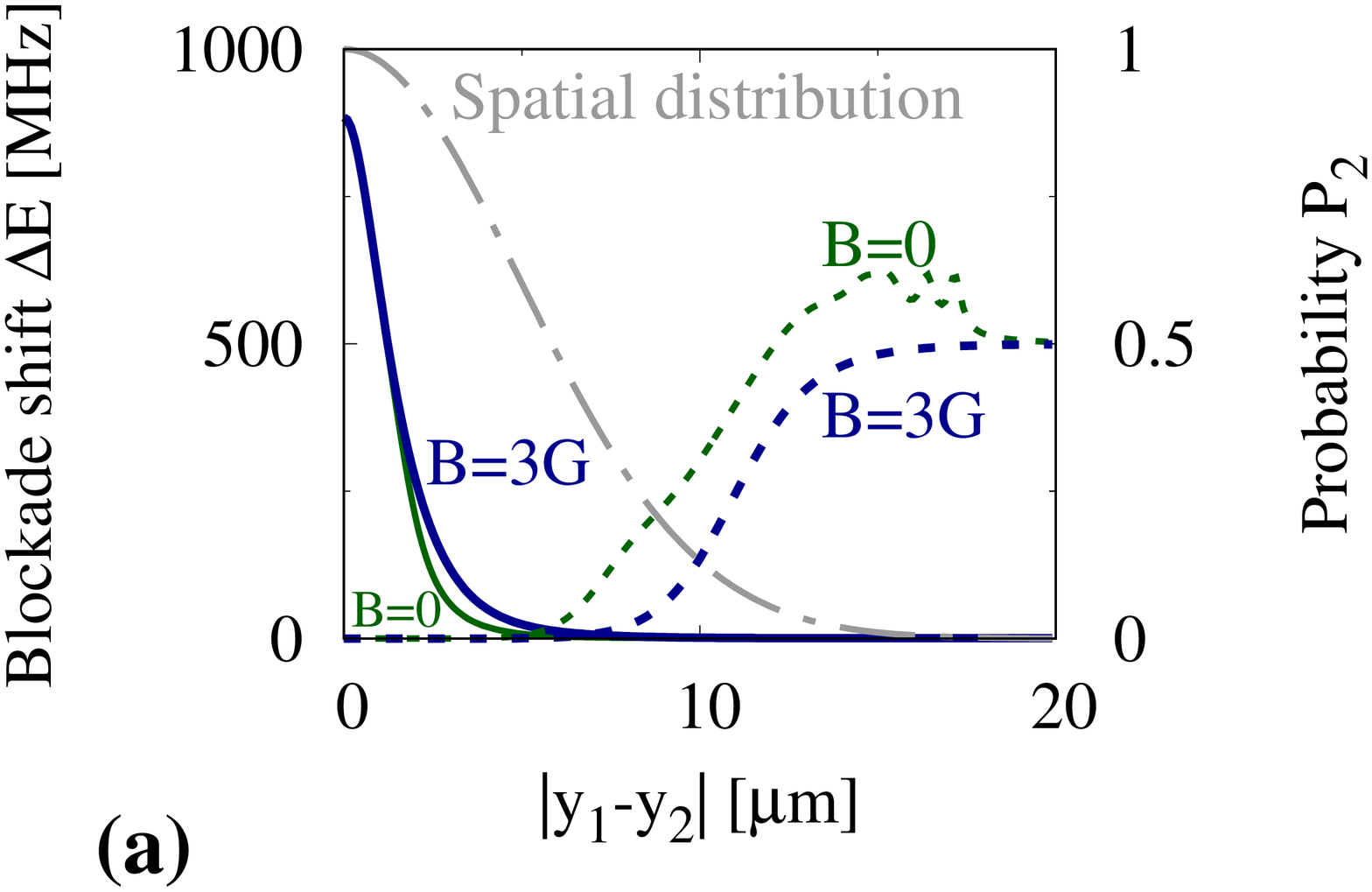}
    \end{minipage}
    \hspace*{1cm}
    \begin{minipage}{.5\textwidth}
      \includegraphics[angle=-90,width=.9\textwidth]
      {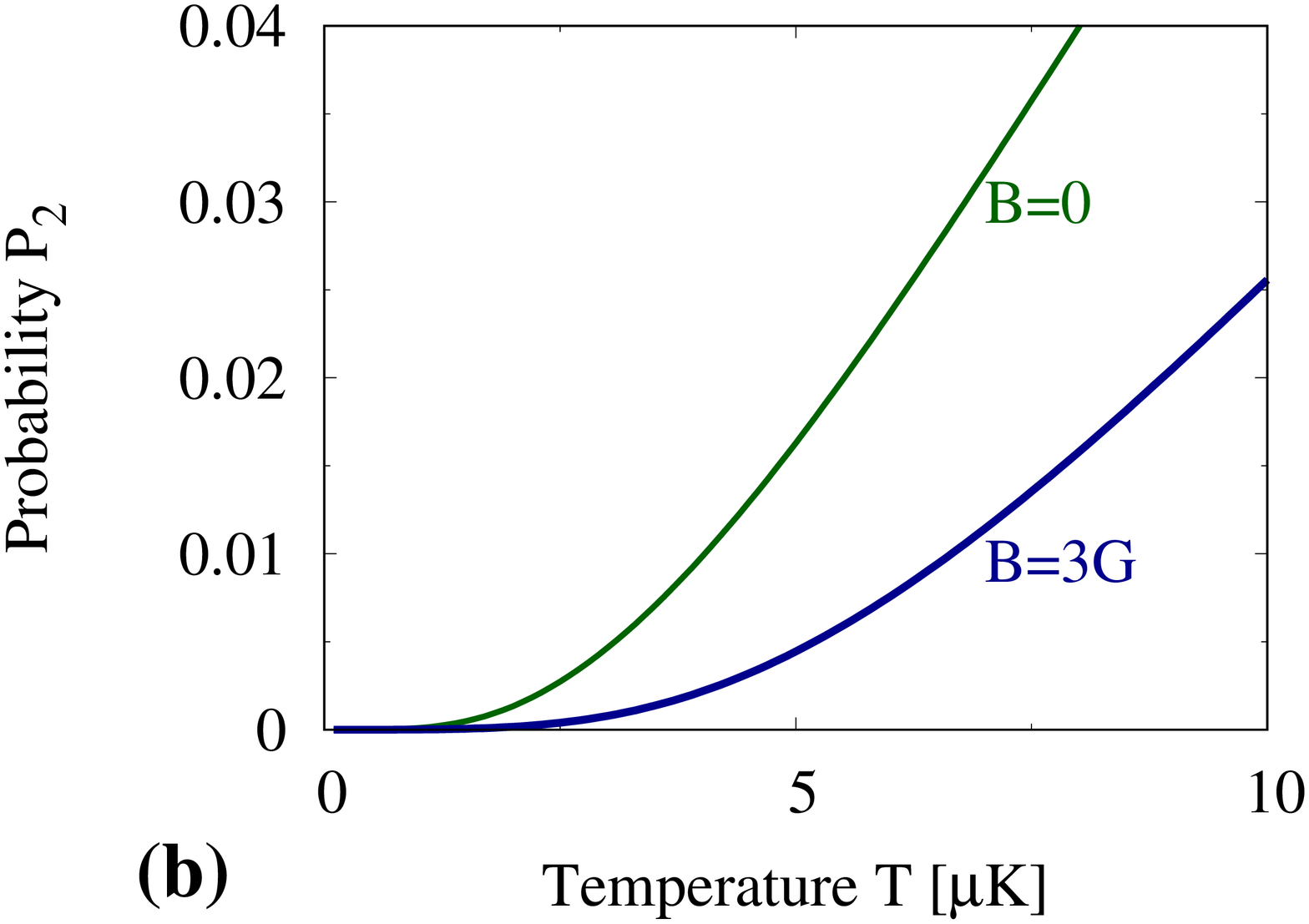}
    \end{minipage}
  %\end{center}
%  \caption{ \label{fig:blockade_theory}
\textbf{Extended Data Figure 2 }
    Calculated heteronuclear Rydberg blockade shift.
     (a) Double--excitation probability $P_{85}$ (dashed curves, right axis)
     and the corresponding blockade shift $\Delta E$
     (solid curves, left axis) as functions of the offset $|y_2-y_1|$.
     The spatial probability distribution
     $p(y=|y_2-y_1|)$, calculated for $T_{87}=8\,\mathrm{\mu K}$ and
     $T_{85}=9\,\mathrm{\mu K}$, is also shown (gray dashed--dotted curve).
     (b) Mean double--excitation probability $<\!P_2\!>$ as a function
     of the mean temperature $T$.
     The thin
     green curve corresponds to $B=0$  and the thick blue curve to
     $B=3\,\mathrm{G}$.
%  }
\end{figure}

\end{methods}

\end{document}